\documentclass[preprint,12pt]{elsarticle}

\usepackage{amssymb}
\usepackage{amsmath}
\usepackage{graphicx}
\usepackage{algpseudocode}
\usepackage{algorithm}
\usepackage{hyperref}

\newcounter{bla}

\journal{Computer Physics Communications}

\begin{document}

\begin{frontmatter}



\title{Identifying Clusters on a Discrete Periodic Lattice via Machine Learning}


\author[a]{Everest Law}

\cortext[author] {Corresponding author.\\\textit{E-mail address:} everestl@usc.edu}
\address[a]{Department of Physics \& Astronomy and Molecular Computational Biology Program, Department of Biological Sciences, University of Southern California, Los Angeles, California 90089, USA}

\begin{abstract}
Given the ubiquity of lattice models in physics, it is imperative for researchers to possess robust methods for quantifying clusters on the lattice --- whether they be Ising spins or clumps of molecules. Inspired by biophysical studies, we present Python code for handling clusters on a 2D periodic lattice. Properties of individual clusters, such as their area, can be obtained with a few function calls. Our code invokes an unsupervised machine learning method called hierarchical clustering, which is simultaneously effective for the present problem and simple enough for non-experts to grasp qualitatively. Moreover, our code transparently merges clusters neighboring each other across periodic boundaries using breadth-first search (BFS), an algorithm well-documented in computer science pedagogy. The fact that our code is written in Python --- instead of proprietary languages --- further enhances its value for reproducible science.
\end{abstract}


\end{frontmatter}



{\bf PROGRAM SUMMARY}

\begin{small}
\noindent
{\em Program Title:} Cluster Collector                                          \\
{\em Licensing provisions:} Creative Common by 4.0                                   \\
{\em Programming language:} Python                                  \\
{\em Nature of problem:}  Lattice simulations of, say, membrane proteins model the spatiotemporal organization of a system. In order to extract insights from such simulations, we need robust methods for identifying clusters of simulated objects on the lattice. \\
{\em Solution method:} Hierarchical clustering first identifies all potential clusters. Then, breadth-first search connects together clusters that neighbor each other across periodic boundaries. \\
\end{small}

\section{Introduction}
There has long been an interplay between physics and subsets of ``machine learning'', before the latter became widely known outside technical fields. For instance, clustering methods has been utilized to speed up force calculations in N-body cosmological simulations \cite{murtagh_hierarchical_1988}. In turn, physics has inspired new algorithms, such as Super Paramagnetic Clustering (SPC), which was applied to image classification problems \cite{domany_superparamagnetic_1999}. In general, clustering methods expose natural classes within data while making as few assumptions as possible. These methods are well-documented by pedagogical publications \cite{james_introduction_2017, hastie_elements_2016} and can be readily adopted by physicists. 

This paper will apply hierarchical clustering to discrete lattice systems in biophysics. These lattices arise when we model the spatiotemporal organization of cell membrane proteins, the study of which would shed light on synaptic transmission \cite{li_stochastic_2017, shomar_cooperative_2017}, viral infections \cite{szklarczyk_receptor_2013}, and inter-cellular communication \cite{noauthor_information_2004} among other areas. In these lattice models, the membrane is discretized into patches according to specific biophysical considerations \cite{gillespie_validity_2014, smith_spatial_2018}, with each patch described by, for instance, a $k-$dimensional vector recording the  amounts of the $k$ chemical species present. Making certain assumptions such as fast diffusion within each lattice site \cite{gillespie_validity_2014, smith_spatial_2018}, we can simulate the system's (stochastic) time evolution using Gillespie's algorithm; see \cite{erban_practical_2007} for a recent review. 

Given the importance of membrane protein organization, it is vital to develop techniques for identifying and describing molecular clusters on a simulation lattice. Hierarchical clustering is one possible choice. To our knowledge, this technique has first been utilized in membrane biophysics by Shomar \emph{et al} \cite{shomar_cooperative_2017}, who in turn were inspired by a nanoscopy experiment \cite{tang_trans-synaptic_2016}. The MATLAB clustering code in \cite{shomar_cooperative_2017} was published as part of its Supplementary Materials. Unfortunately, that code does not model diffusion, and therefore does not correspond to any particular boundary conditions.

Here we present an improved implementation which may be useful to biophysicists and practitioners of other fields alike. First and foremost, assuming a diffusive system, our code readily connects together lattice sites that touch across periodic boundaries. Given the prevalence of periodic boundary conditions, we believe this is an important feature. Second, our code is written in Python, which unlike MATLAB is non-proprietary and free-to-use. As Python becomes one of the most popular programming languages globally \footnote{See for example the 2018 survey on Stack Overflow, a leading forum for exchanging ideas on software development. \texttt{https://insights.stackoverflow.com/survey/2018/}}, we believe that research code written in Python is easier to maintain and adopt for other purposes, thus leading to more reproducible scientific results. 

This paper is structured as follows. Section 2 gives an overview of the hierarchical clustering method. Section 3 describes the 2D lattices to which our code is applied, and how periodic boundaries are handled. A summary and conclusion follow in Section 4. All code can be found online at \url{https://github.com/openerror/PhysicsLatticeClustering}. 

\section{Hierarchical Clustering --- Theory and Example}
\subsection{Overview of Algorithm}
There are many implementations of hierarchical clustering \cite{verma_effective_2017, noauthor_fastcluster:_nodate}, some of which are parallelized \cite{dash_efficient_2004, olson_parallel_1995, dahlhaus_parallel_2000}. In this paper we will work with the \emph{agglomerative} variant, which forms clusters from the bottom-up --- i.e. starting from single observations \cite{hastie_elements_2016}. Below we will give a qualitative overview of the procedure, assuming serial computations, followed by a toy example. 

Assume that we have $N$ observations in a $D$ dimensional space. To begin, we need a ``dissimilarity measure'' for quantifying the difference between observations; Euclidean distance is a common choice, and a natural one when considering physical separations. With a dissimilarity measure defined, the algorithm iteratively merges the two observations or \emph{clusters} that are the most similar. After $N - 1$ steps we obtain a single ``megacluster'' containing all the original $N$ observations, and the algorithm terminates.

To make concrete the dissimilarity $d$ between clusters of observations, various schemes or \emph{linkages} have been developed. For this work we have adopted ``single'' linkage \cite{hastie_elements_2016}: 
\begin{align}
d(u,v) = \textnormal{min} (dist(u[i], v[j])
\end{align}
for all points $i$ in cluster $u$ and $j$ in cluster $v$. In other words, we compute all pairwise dissimilarities between points from $u$ and $v$, take the minimum, and let that be the dissimilarity between the two clusters.

The next and final step is to extract clusters at a desired scale from the ``megacluster'' produced. It involves retaining all clusters that are merged at a dissimilarity lower than some chosen threshold. To make this more concrete, we present a toy example below.

\subsection{An Example}
\begin{figure}
\center
\includegraphics[width=0.8\textwidth]{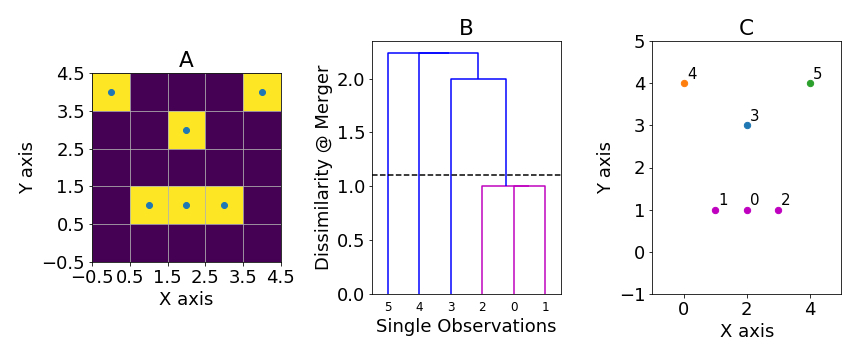}
\caption{Sample data showcasing hierarchical clustering with single linkage. A) The 5x5 lattice on which we ran the clustering algorithm; yellow squares indicate ``occupied'' sites, and touching sites are considered part the same cluster. B) Dendrogram tree demonstrating mergers of observations and clusters throughout the procedure. Leaf nodes indicate single observations, which are merged at nodes higher in the graph; the vertical height at which nodes are located represent the dissimilarity between the pair of merged clusters/observations. C) The six observations, labeled with the same numbers as the dendrogram leaves.} 
\label{fig:sample}
\end{figure}

Say we have a 5-by-5 lattice with 6 occupied sites: $(x,y) = (0,4), (1,1), (2,3), \\(2,1), (3,1), (4,4)$. (See Figure \ref{fig:sample}.) Assume that sites neighboring each other belong to the same cluster, and that only non-diagonal neighbors are considered. With the stated criteria, we expect to obtain 4 clusters.

We can visualize the merger of observations and clusters through a tree-like ``dendrogram''. 
The leaves nodes at the bottom of the dendrogram represent the original $N$ observations. As we move up the graph, we see the leaves merging into new nodes, which may undergo further mergers. The height in the dendrogram, at which a pair of nodes combine, indicate the dissimilarity between the nodes. 

The dendrogram in Figure \ref{fig:sample}B is consistent with the hierarchical clustering described. Going from the bottom to the top, we see that the sites 0, 1, 2, which are neighboring each other, are first merged together. The 3-member cluster produced is then merged with the sites at the top left and right corners successively.

In order to obtain clusters at the desired scale, we cut the dendrogram horizontally. Any mergers performed above the cut are ignored, while those that occur at one level below are retained. For our particular example, any cut between unity and $\sqrt{2}$ would work, and allow us to identify sites that touch each other through their ``top'', ``bottom'', ``left'' and ``right''. In the end, we obtain four clusters --- as expected from the definition of clusters laid out in the beginning of Section 2.

\subsection{Sample Code and Required Libraries}
For the code that produced Figure \ref{fig:sample}, please see the Jupyter notebook submitted with this publication. All functionality presented so far is implemented in the function \texttt{detectClusters}, located within the file \texttt{clustering.py}. Please see the docstring under the function declaration for details.

All required Python libraries are listed in \texttt{requirements.txt}; the hierarchical clustering routines in particular are shipped with the \texttt{SciPy}.  Given an existing Python installation, these libraries can be installed quickly using the console command. 
\begin{align*}
\textnormal{\texttt{pip install -r requirements.txt}}
\end{align*}
or through whichever package manager preferred. 

\section{Handling Periodic Boundaries of the 2D Lattice}
When simulating large systems, periodic boundaries are commonly adopted to make computations more tractable. How then can we take into account periodic boundaries in hierarchical clustering? While it would be costly to modify an existing implementation directly, we can achieve our goal by processing the simulation lattice and the clustering output together. The logic is as follows: 
\begin{enumerate}
\item Identify clusters that are touching the boundary, and determine the coordinates where the touching occurs.
\item For the clusters identified, check if they neighbor each other across the periodic boundaries.
\item If they do, consider the touching clusters as one.
\end{enumerate}

The above logic is implemented in the functions \texttt{extractClusterCoordinates()} and \texttt{grouping()}, located within \texttt{clustering.py} and \texttt{grouping.py} respectively. The former function takes the output of \texttt{detectClusters()}, and determines which clusters touch the boundaries. The data thus processed is then sent into \texttt{grouping()} to generate the final output. The function \texttt{grouping()} returns a Python list containing integer tuples, each of which describes a cluster --- merged across periodic boundaries or otherwise. 

The integers themselves correspond to each individual cluster identified and returned by \texttt{detectClusters()}. For identical input data, SciPy's hierarchical clustering implementation would give identical cluster IDs. For instance, the system depicted in Figure \ref{fig:sample} corresponds to the output \texttt{[(0), (1), (2,3)]}; 0 corresponds to the 3-member cluster, and (2,3) are the two observations on the top left and right corners, which would be merged across periodic boundaries.

\subsection{The Use of Breadth-First Search To Merge Clusters}
One non-trivial task that our code has to accomplish is to identify clusters that are touching ``transitively''. For instance, clusters A and C may be neighbors of cluster B, but they do not touch directly. How then can we robustly identify all three of them as one single cluster?

To accomplish the task above we have utilized Breadth-First Search (BFS), a standard technique in computer science. Given a graph comprised of \emph{nodes} and (undirected) \emph{edges}, starting from a given node BFS would find all other nodes reachable by traversing an edge\footnote{For an illustration, see the first 1:40 of https://www.youtube.com/watch?v=0u78hx-66Xk}. 

\begin{figure}[h]
\center
\includegraphics[width=0.8\textwidth]{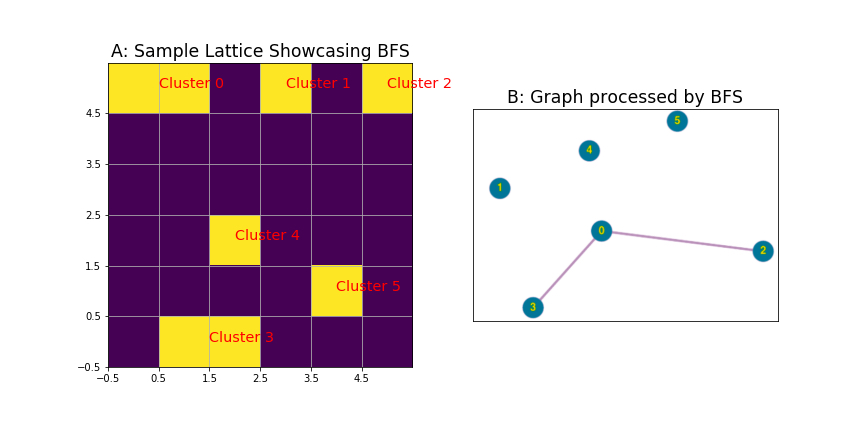}
\caption{Using BFS to identify cluster neighbors from across periodic boundaries. A) The lattice on which we ran the clustering routines. Clusters 2 and 3 are not neighbors themselves, but they both touch cluster 0 across periodic boundaries. Therefore all three together are considered as one cluster. B) Graph representing the connectivities of the clusters in A. Applying BFS onto this graph, represented as a matrix $C$ in computations, gives us the connected pieces: (0, 2, 3), (1), (4), (5).} 
\label{fig:bfs}
\end{figure}

The problem of merging transitively-neighboring clusters can be translated into a graph-search problem: clusters become ``nodes'' that connect to their direct neighbors through ``edges''. Given the direct connectivities between each pair of clusters, we can form a square \emph{connectivity matrix} $C$, whose height is the same as the number of clusters. Each entry $C_{ij} = 1$ indicates an ``edge'' between clusters $i$ and $j$; $C_{ij} = 0$ otherwise. For the lattice presented in Figure 2, the corresponding matrix $C$ is

\begin{align*}
\begin{bmatrix}
0 & 0 & 1 & 1 & 0 & 0 \\
0 & 0 & 0 & 0 & 0 & 0\\
1 & 0 & 0 & 0 & 0 & 0\\
1 & 0 & 0 & 0 & 0 & 0\\
0 & 0 & 0 & 0 & 0 & 0\\
0 & 0 & 0 & 0 & 0 & 0\\
\end{bmatrix}.
\end{align*}

Since clusters touching do not involve directions, the BFS algorithm is working with an undirected graph, which leads to a symmetric matrix $C$. Starting from any given cluster or ``node'', BFS would find all other ``nodes'' connected to it; repeat for all nodes and obtain the desired output.

Implementation-wise, the function \texttt{connectedComponents()} in \texttt{BFS.py} takes as input a matrix $C$; it starts at each (unvisited) cluster, and finds all others connected to it. Finally it returns the groups of connected clusters as a Python list. This is in fact what \texttt{grouping()} does, \emph{after} processing its own input to form the matrix $C$. 

For clarity, we present below pseudocode describing the entire procedure depicted in Sections 2 and 3.
\begin{algorithm}
\begin{algorithmic}
\State array $\gets$ 2D simulation lattice
\If {array not binarized}
	\State binarize(array) \Comment i.e. turns into 1s and 0s, with a thresholding scheme
\EndIf
\State
\State HCStats $\gets$ detectCluster(array) \Comment Performs hierarchical clustering
\State \State \Comment Determine which clusters touch the boundaries
\State lattice$\_$params $\gets$ \{ `sizeY': lattice height, `sizeX': lattice width\}
\State cluster\_coordinates $\gets$ extractClusterCoordinates(HCStats, \\lattice$\_$params)

\State \State \Comment Obtain tuples of integer clusters IDs
\State cluster\_groups $\gets$ grouping(HCStats)
\end{algorithmic}
\end{algorithm}

\section{Conclusions}
To sum up, this paper has solved a common problem in lattice simulations, using only a combination of standard techniques from machine learning and computer science. All algorithms used --- namely, hierarchical clustering and breadth-first search --- are well-documented, and optimized implementations of clustering are easily available as pre-built libraries. We hope that our work would inspire further imports of information-technological techniques into physics.

Although this paper has only worked on 2D lattices, with a few modifications the same code can be applied to a rectangular 3D lattice. However, for more complicated geometries substantial edits would be necessary. In particular, we may need a new measure for the dissimilarity between observations, in addition to alternative methods for handling boundary points or even non-periodic boundary conditions. Non-trivial geometries arise in attempts to accurately describe molecular diffusion on cellular membranes \cite{fange_mesord_2012, roberts_lattice_2013}, and present additional needs for appropriate analytical techniques.

\section{Acknowledgements}
This work was supported by NSF award number DMR-1554716 and the USC Center for High-Performance Computing. Declarations of interests: none.
\section*{References}





\bibliographystyle{elsarticle-num}
\bibliography{CPiPRef}






\end{document}